# CONTEXT AWARE MOBILE INITIATED HANDOFF FOR PERFORMANCE IMPROVEMENT IN IEEE 802.11 NETWORKS

Abhijit Sarma, Shantanu Joshi and Sukumar Nandi

`abhijit_gu@yahoo.com`, `s.joshi@iitg.ac.in`, `sukumar@iitg.ac.in`

Department of Computer Science and Engineering,
Indian Institute of Technology,
Guwahati, India

## ABSTRACT

*IEEE 802.11 is a widely used wireless LAN standard which offers a good bandwidth at low cost In an ESS, multiple APs can co-exist with overlapping coverage area. A mobile node connects to the AP from which it receives the best signal. Changes in traffic to and from different MNs occur over time. Load imbalance may develop on different APs. Throughput and delay of the different flows passing through the APs, where the load has increased beyond certain limit, may degrade. Different MNs associated to the overloaded APs will experience performance degradation. Overall performance of the ESS will also drop. In this paper we propose a scheme where MNs experiencing degraded performance will initiate action and with assistance from the associate AP perform handoff to less loaded AP within its range to improve performance.*

## 1. INTRODUCTION

Mobile wireless communication is gaining popularity day by day. IEEE 802.11 [1] is the most popular standard wireless local area network (WLAN) in which a mobile user can connect to a local area network (LAN) through a wireless connection. It's popularity is due to it's low cost and ease of deployment [2][3]. In a WLAN mltiple access points (AP) may co-exists with overlapping coverage area. To increase the total available bandwidth and to allow more mobile nodes (MN) to connect, deployment of more than one interconnected AP with overlapping coverage area is usual. A 802.11 WLAN with interconnected APs forms an Extended Service Set (ESS). Deployment of a WLAN is normally limitted to buildings and campuses. To serve a larger coverage area, a WLAN may be connected to a wireless metropolitan area network (WMAN). User moving out of the coverage area of WLAN will be able to communicate through the WMAN. IEEE 802.16 is a popular WMAN standard. An interconnected system of WLAN ESS and WMAN will provide greater mobility to users. In an ESS, a MN will connect to the AP from which the MN receives the best signal disregard of the existing load on the different APs within its range. This may lead to an imbalance in the load on the different APs in the ESS. In[4], [5] amd [6] it is shown that traffic load is often unevenly distributed across APs. In [7] it is shown that WLAN attains maximum throughput and minimum latency when the nodes are unsaturated. WLAN users tend to concentrate in some areas [5]. AP nearest to such an area is expected to get overloaded. Decrease in performance of the different flows will occur when some APs gets overloaded even if some other APs within the same ESS have spare bandwidth. The applications in the MNs associated to such overloaded APs will experience reduced quality of service (QoS) in terms of throughput, delay and packet loss. The overall throughput of the ESS also suffers. As the load offered to an AP may change dynamically even when there is no change in the number of MNs associated, admission control is not effective in solving this problem. The QoS of the suffering Aps and the overall throughput of the ESS may be increased if some load of the overloaded APs can be transferred to the less loaded APs by handing off some of the associated MNs from the overloaded APs to the less loaded APs. When the load of the ESS increases beyond a limit, the performance of the whole ESS starts to suffer. in an interconnected WLAN-WMAN system, a MN will prefer to remain connected to WLAN because of cost considerations. However, when MNs starts suffering from performance degradation due to overload in the ESS, it may be transferred to connect to the base station(BS) of a WMAN. In this paper we propose a scheme where the mobile nodes experiencing QoS degradation tries, with the assistance of its associated AP, to handoff to some alternate APs. The MNs experiencing performance loss tries to find alternate APs for handoff. As the MN does not know the load conditions on these APs it does not handoff to such an AP immediately but sends the list of such APs along with its own load information to the associated AP. The associated AP exchanges messages with the given APs to know the load on those APs and based on that, it selects a list of APs to which the MN can handoff and communicates the same to the MN. The MN chooses one of such APs and performs handoff. In this way, loads move from overloaded APs to the less loaded APs. Due to this, the bandwidth is better shared among the APs within an ESS and the overall throughput of the ESS increases. In case the suffering MN does not find a suitable AP to handoff, it performs vertical handoff to a BS covering the area.

## 1.1. Our Contribution

Our main contribution is to increase throughput to meet the offered traffic in an ESS with minimal load adjustment. The following are the salient features of our scheme : -

- Load adjusted only when MNs suffers

- MNs provide the list of APs which are reachable from the MN.

- AP offers alternatives to MN based on load on the MN and load on other APs as provided by MN.

- MN takes the final decision to shift association based on alternatives offered by AP.

- Minimal numbers of MNs changing association.

- APs communicated via infrastructure, MNs and wireless channels are not involved.

- Very low scanning overhead to locate alternate APs.

- Perfect balance is not tried. Balancing only enough to maximize throughput.

- MN perform handoff to WMAN when all covering APs in ESS gets overloaded.

The rest of the paper is organized as follows. Section 2 introduces the IEEE 802.11 ESS. In section 3 we have a brief introduction to earlier works targeted to performance improvement in IEEE 802.11 WLAN by distributing load among APs. In section 4 we describe our scheme. The simulation results are provided in section 5. In section 6 we conclude.

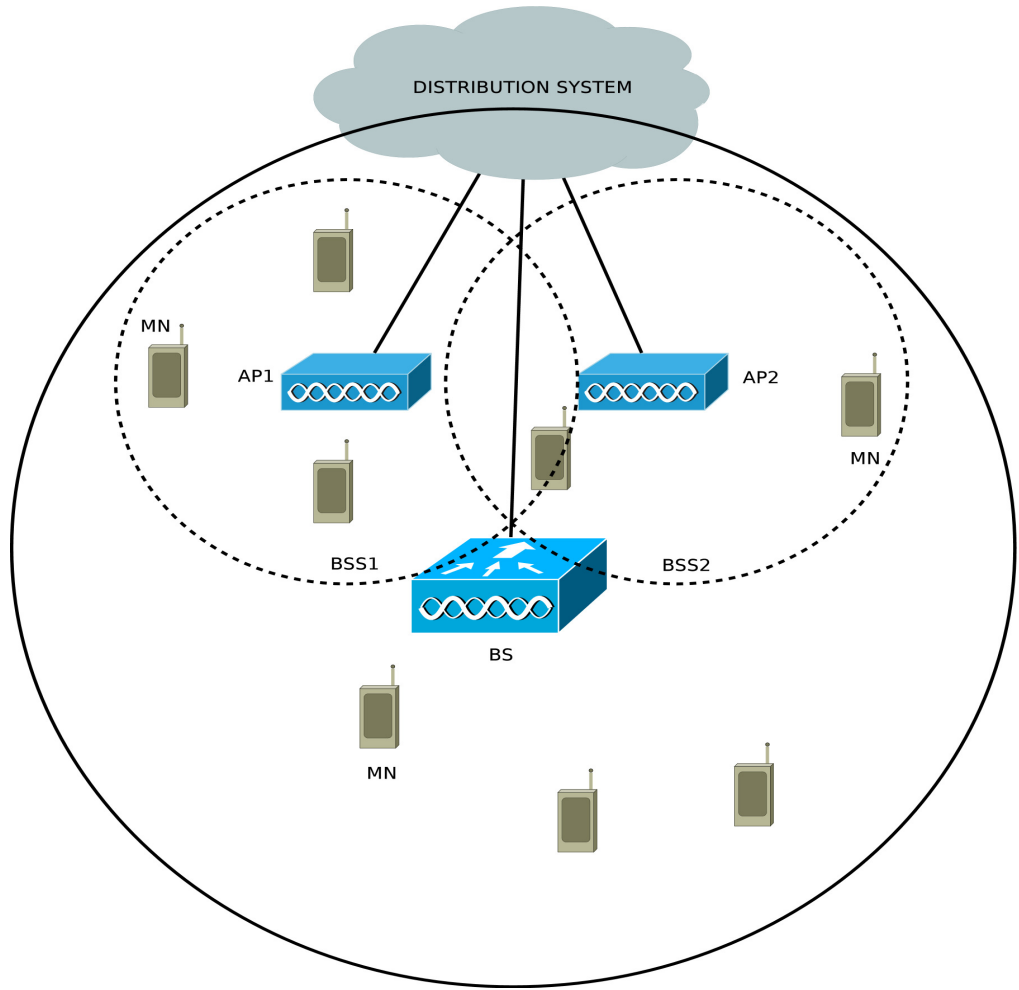

Figure 1. A WiFi-WiMax heterogenous network.

## 2. BACKGROUND

In IEEE 802.11 WLAN [1], there can be two types of nodes present, namely AP and MN. Of these, APs usually have fixed location while MNs are roaming devices. A basic service set (BSS) is a set of stations that communicate with one another.

BSS can operate in two modes. In the independent BSS (IBSS) mode there is no connection to a wired network. IBSS is a temporary netowrk setup for a perticular purpose. We shall not refer to this mode any further in this paper. The other mode, called an infrastructure BSS, is characterized by the presence of an AP. In this mode all communication among the mobile stations is performed through the AP. An extended service set (ESS) is a set of infrastructure BSSs, where the APs communicate among themselves to forward traffic from one BSS to another as well as to and from other networks. It also facilitate movement of MNs between BSSs. To avoid disruption of communication to MNs, different APs should have overlapped coverage area. There can be up to fourteen different channels in the 2.4 GHz Industrial, Scientific and Medical (ISM) band in which a 802.11 WLAN may operate. Due to restrictions imposed, not all channels are available in all countries. Channels one to eleven are

International Journal of Computer Networks & Communications (IJCNC) Vol.3, No.3, May 2011

available in most of the countries. Each channel is of 22 MHz bandwidth but spaced only 5 MHz apart. As such there can be only three non overlapping channels in the eleven channels. Although the APs may operate on overlapping channels, it is not desirable to use overlapping channels in adjacent APs with overlapping coverage area due to interference and resulting performance loss. To increase the available bandwidth in an ESS the APs in an ESS should operate in different non overlapping channels. A MN should associate to the AP of a BSS to communicate. A MN associates to an AP by sending an association request message. The AP, if willing to allow association, replies with an association response message and the MN gets associated to the AP. At the time of associating an MN, the AP considers only its own status to decide whether the MN will be allowed association or not. The MN may be within the coverage area of another AP with lesser load and association to such an AP might have given a better distribution of load. As such, considerable load imbalance among the different APs in an ESS may occur resulting in degradation of performance of the ESS.

In a 802.16 WMAN [8], there can be two types of nodes, namely, subscriber station (SS) and base station (BS). BSs have fixed location where SSs may be mobile. All communications to and from a SS must go through BS. BSs may be connected through backbone and also to another network. In 802.16 WMAN, a MN needs to perform network entry with the BS before it can communicate through the BS. To form a heterogeneous network of WiFi and WiMax networks, these networks may be connected togather by interconnecting the BSs and APs through a backbone. This backbone is usually a wired network. Such a setup is shown in Figure 1.

## 3. RELATED WORKS

Most of the existing works target to performance improvement in terms of delay and throughput in 802.11 multi AP networks are centered around balancing load in the different APs. There are several such works. Gilbert Sawma et. el. presented a scheme in [9] where the loads in the APs are adjusted dynamically whenever a new MN tries to associate to an AP and the AP is already overloaded. In their scheme, several MNs are directed to change AP to balance the load. In their scheme, balancing is done only when a new MN is associated to the WLAN but if load changes due to increase in load from already associated AP no balancing is done. Moreover, the decision to switch APs is made solely by the AP based on the load condition of APs only. In the scheme presented in [10] a periodic probing is used in each MN which requires about 300 ms per probe. This recurring disruption in communication is unacceptable for real time traffic. The scheme also takes only AP load into account. In [11] also the load of only AP is considered. In [12] the authors are considering RSSI only for load balancing. However, RSSI can not be a good indicator of load. In [13] the AP decides which MNs should handoff. Also, the nearby access points are found by the AP. In [14] a static assignment of channels an placement of AP is done. It does not cater for the dynamic changes in the load pattern. In [15],[16], [17] and [18] the decision is taken centrally and MNs are directed to move all at once. In [19] the decision is solely based on MN. AP broadcasts its load information through beacons. In [18] cell breathing technique as used in cellular mobiles is used. In this technique, the MNs are forced to handoff irrespective of their load condition and desire. In [20] a decision is taken centrally in an AP. In this scheme the throughput per AP is used as a metric for load.  In [17] The authors use wireless medium busy time (MBT) as the metric from load.

There are a few schemes proposed for performance improvement in WiFi-WiMax Heterogeneous network. In [21], the authors have proposed a scheme which requires load information of different networks which is difficult to gather in a network withn degrading performance. The scheme proposes two criteria for handoff between WLAN and WMAN. In one case they proposed RSSI as the criteria and in the other case they proposed the ratio of load between different networks as the criteria. In [22] the authors proposed a scheme where MNs

51

perform handoff whenever the QoS of applications degrade. The exact mechanism of how it is achieved is not discussed in the paper. In [23], the authors used packet loss and delay as critera for performing handoff from one network to another. The authors reported 9 ms maximum delay which is not possible in WiMax without altering default parameters. In [24], authors proposed an integration model for Wi-Fi and WiMax network and BS assisted scheduling algorithm to provide QoS guarantee. In order to integrate Wi-Fi and WiMax at MAC layers authors provided mapping between access categories in WiFi and sevice classes in WiMax. A scheduling algorithm is proposed in their scheme for providing QoS guarantee for real-time traffic in terms of delay and non real-time traffic in terms of packet loss. In [25], the authors proposed a decision making process for mobile devices to select one network from among several available options in the surroundings. Mobile node gathers the parameters reflecting network performance as well as its current location. This information is then used for taking handoff decision. The drawback of this scheme is that it does not specify details of which parameters are measured, how the parameters are measured and collected. For the MN to handoff, it needs to find an alternate AP. The above schemes assumes that the MN performs a full scan to locate an alternate AP. In [26] the authors uses a interleaved scanning technique to reduce handoff time in WLAN. A modified version of this technique is also used in load balancing schemes in [27] and [28].

## 4. THE PROPOSED SCHEME

In this paper we propose a scheme where MNs which experience degradation of performance are handed over to other AP in th ESS if an underloaded AP is found. In case an underloaded AP is not found within the radio range of the MN, the MN is ahnded off to a covering BS. This improves the throughput and delay of the suffering MNs and also the ESS.

### 4.1. Assumptions

The following assumptions are made: -

- APs and BS are connected through a backbone.
- Coverage area of ESS falls within the coverage area of BS.
- BS has sufficient spare capacity.
- APs have overlapping coverage area.
- Overlapping BSSs operate in non overlapping channels.

### 4.2. Details of the Scheme

Whenever a MN senses a degradation of performance, it scans the different channels to find suitable APs to which it can perform handoff. The degradation of performance is detected using the methods which are presented in subsection 4.3 To avoid the long scanning latency, the MN scans using a technique which is explained in subsection 4.4. The MN thus collects a list of APs with which it can associate if needed. Let the list of such APs be called *APList*. The MN then informs its associated AP, say $AP_a$ about its desire to disassociate from $AP_a$ and re associate with another AP where a better performance may be expected. To do this, the MN sends a *MoveRequest* message containing the *APList* and its desired throughput to $AP_a$. The MN keeps updating the exponentially weighted moving average (EWMA)[29] of the queue length and sends the *MoveRequest* message repeatedly at a fixed interval of $t_{repeat}$, for upto $n_{repeat}$ times or until it gets a response from $AP_a$. Whenever $AP_a$ receives a *MoveRequest* message from a MN, it sends a *LoadRequest* message to all the APs in the *APList* through the backbone. The APs receiving the *LoadReques* message, replies with a *LoadResponse* message through the backbone containing its current load status. $AP_a$ receives the *LoadReesponse* messages within a given time and based on the replies it receives and the APList it received from the MN, selects zero or more APs to which the MN can handoff. $AP_a$ keeps the list of such APs in *HCList*. Let $AP_i$ be the $i^{th}$ AP in the ESS. Let the concerned MN be associated to the $j^{th}$ AP. Let $L_a$ be the traffic passing through the $j^{th}$ AP. Let $L_i$ be the traffic passing through the $i^{th}$ AP and $M$ is the load on the MN. Let also the set of APs in the *HCList* be *HC*. Then: -

$$HC = \{AP_i \mid i \neq j \text{ and } L_a - M - L_i > \Delta\} \quad (1)$$

Where $L_i$, $M$, $L_a$ and $\Delta$ are traffic volume in bits per second. $\Delta$ is a small fraction of the maximum throughput. The value of $\Delta$ is chosen considering the fact that, if it is too small, the MN that performed handoff to get better performance will tend to handoff again and too many handoffs may occur unnecessarily. If it is too large, potential opportunity to increase performance may be lost. $AP_a$ sends the *HandoffCandidateList* in its reply message, *HandoffTargetMessage*, to the MN. The MN, based on the signal to interference plus noise ratio (SINR), selects one of the APs in this list and performs handoff to that AP. The MN selects the AP from which it receives the strongest SINR. It is likely that more than one MN sends *MoveRequest* messages to $AP_a$. If all such MNs are allowed to handoff to another AP, a severe imbalance in load and hence degradation of performance may result. To prevent it, whenever $AP_a$ processes a *MoveRequest* message, it drops all further *MoveRequest* from any MN for some fix period of time Tignore. It may so happen that no suitable AP could be found to which the MN can perform handoff. In that case it is proposed that the MN find a suitable BS to handoff. To do that, whebever the EWMA of queue length crosses a certain limit, the MN activates it's WiMax interface and starts scanning for BS. As the curent communication still continues through the WLAN interface, there is no disruption of communication. However, based on the overload on the associated AP, there may be some performance degradation for a brief moment till the MN performs handoff. When a suitable BS is found, the MN records the same. After getting the *HandoffCandidateList*, if the MN observes that the list is empty, it performs network entry with the BS found earlier earlier and transfers all flows to the BS. It does so while still communicating through the associated AP. After all the flows are successfully transfered, the MN sends a route update (RU) messageas through the BS. It then disassociates from the associated AP and starts communicating through the BS. While connected through the BS, the MN periodically scanns for WLAN using interleave scanning technique. If an AP is found, MN sends a *LoadRequestFromMN* to the AP. The AP replies with a *LoadResponse* message where it informs about its spare load. If sufficient spare load exists in the AP, the MN associates with the AP, sends RU message through the AP and disconnects from th BS. The destination of the RU message sent by the MN is the correspondant node (CN) the MN is communicating. Each node receiving RU message will update their routing table to reflect the current location of the MN. This is required so that when the MN is changing the network, the routers along the path from MN to CN update their routing tables quickly. In absent of the RU message, the route will be updated by the routers after some time based on the routing protocol used. This time may be considerably large. Communication will be disrupted during the time from handoff till all the routers update their routing tables. This will also have a significant effect on tcp congestion window which will cause more end to end delay. Real time traffic will suffer badly.

### 4.3. Detection of Performance Degradation

The degradation of performance is sensed in two different ways. In the first method, the packet drop rate is used to sense the degradation of performance. A moving average of the number of packet dropped within a time window is kept. In this whenever the average increases above some threshold, the MN tries to find an alternate AP for association. It is observed that, when the traffic in the ESS increases gradually, packets gets delayed considerably even when the throughput does not get effected significantly. So, changing the association of MNs when degradation of performance is sensed this way will result in improvement of performance in terms of throughput but there may be still degradation of performance in terms of delay. In the second method the degradation of performance is sensed by observing the interface queue length. To smooth out the variations in the interface queue length the EWMA of the queue length is calculated every $t$ units of time. Let $E_t$ be the EWMA of the queue length at time $t$. Let $E_{t-1}$ be the EWMA of the queue length at time $t-1$ and let $Y_{t-1}$ be the actual queue length at time $t-1$. Then we calculate $E_t$ as

$$E_t = a \times Y_{t-1} + a \times E_{t-1}, \ 0 < a \leq 1, \ t > 3 \qquad (2)$$

The value of a decides how fast the MN forgets the old values of *Y*. Whenever $E_t$ crosses a threshold, the MN tries to find an alternate AP for association.

### 4.4. Locating an AP to Handoff

To avoid disruption of communication while locating a candidate AP for handoff the MN avoids a full scan of all channels which is usually done. Instead, the MN tries to scan one channel at a time. Data communication continues in between scans for each channel. To scan a channel, the MN switches to that channel and sends a *DSProbe* message in that channel and immediately comes back to the channel in which it was communicating. This message contains the MNs address and also the associated AP's address. The APs, if any, receiving the *DSProbe* message sends a *DSProbeResponse* message to the associated AP through the distribution system (DS). The associated AP, after receiving the *DSProbeResponse* messages forwards the mesage to the MN. Thus the MN finds the diferent APs within it's radio range. MN keeps the information of different APs found during scanning in a list. Let the list be *APList*. When the MN senses degradation of performance, it prepares a circular list of channel numbers to scan. These are the channels that contains all the 802.11 channels except the current channel and the channels that overlaps with the current channel. It initializes a variable *NextChannel* to the channel number of the first channel in the list. It empties *APList*. It also starts a timer. We shall refer this timer as *ScanTimer*. Let the time out value of this timer be *T*. At the expiry of time *T* when the timer event is triggered the MN does the following

- it empties *APList* where the channel number of the AP equals *NextChannel*.

- it changes channel to *NextChannel*.

- it sends a *DSProbe* message and waits for a time period of $t_1$.

- AP receiving the *DSProbe* message sends a *DSProbeResponse* message to the associated AP.

- associated AP forwards the *DSProbeResponse* message to MN.

- if any *DSProbeResponse* the AP information obtained are put into *APList*.

- it changes channel to the channel in which it was communicating with its associated AP.

- it updates *NextChannel* to the entry which is cyclically next to the current value of *NextChannel* in the channel list.

- it starts *ScanTimer* again and starts normal communication.

As the *ScanTimer* is restarted, it will be triggered again and the whole process will repeat again. Till that time normal communication continues. The timer is canceled when all the channels in the circular list of channels are scanned.

### 4.5. Choice of Load Metric

Load metric is an important aspect of a load balancing system. There may be several choices for measuring the load. Global System for Mobile Communications (GSM) uses number of calls to measure load. In GSM each call takes equal amount of bit rate and so number of call is a good metric in GSM. In WLAN, different flows require different bit rates. In [30]} the authors show that metric based on measured traffic is a good metric for load balancing in WLAN. In [31], Number of competing stations is used as metric for load. In [17] the authors use wireless medium busy time (MBT) as the metric. In [32], the authors define client utilization estimate (CUE) as the fraction of time per time unit needed to transmit the flow over the network. CUE is an indicator of network resource usage. CUE is also used in [9] as load metric. In our scheme, we use traffic as load metric. However, for the initiation of the load adjustment process, in the MNs, we use packet drop and EWMA of the interface queue length to sense degradation of performance. Packet drop and EWMA of queue length gives an early indication of performance degradation.

## 5. SIMULATION RESULTS

The proposed handoff scheme is simulated using ns-3 (version 3.8)[33] network simulator. We have modified ns-3 to implement the proposed scheme. In this section we present simulation results to demonstrate the performance of our scheme. Using the experiment, the initialization process and performance of the ESS with respect to throughput and delay are verified. The AP selection procedure is not verified.

### 5.1 Experiment Setup

As shown in Figure 2, two BSSs, namely BSS1 and BSS2 are setup. Access points AP1 and AP2 are placed in BSS1 and BSS2 respectively. Fifteen MNs are placed in such a way that it falls in the coverage area of both BSS1 and BSS2. The dotted circles centered at the Aps are the coverage area of therespective APs. AP1 and AP2 and a correspondent node CN1 are connected to the access router AR1 through point to point links each having a bandwidth of 100Mbps and delay 2 ms.The BSSs operate at different non overlapping channels in the 2.4 GHz ISM band and operate using the IEEE 802.11b standard. All the MNs initially operate in the channel of BSS1 and are associated to AP1. Constant bit rate (CBR) traffic are setup to flow from each MN to CN1. We kept the the default RTS-CTS threshold of ns3 which is set at 1500. Each MN sends packets of size 1500 bytes every 20 ms. The $i^{th}$ MN starts its traffic at the $i^{th}$ second. The initiation of the load adjustment is done by MNs based on rise in packet drop and rise in queue length. A BS is placed in such a way that the coverage area of the BS encompasses the coverage area of both the AP. The BS is connected to the AR through a point to point link having a bandwidth of 10 Mbps and delay 2 ms. Six cases are studied. In cases I, II and III, MNs are not allowed to handoff to BS even when the ESS is overloaded. Throughputs of each AP are measured at each second. The packet delay, that is the time the packet takes from transmission of the packet by the UDP client generating the CBR traffic, to the reception of the packet by the UDP server at the destination, is also measured for observing the performance of the scheme. The value of a as in Equation (2). is set to 0.1.

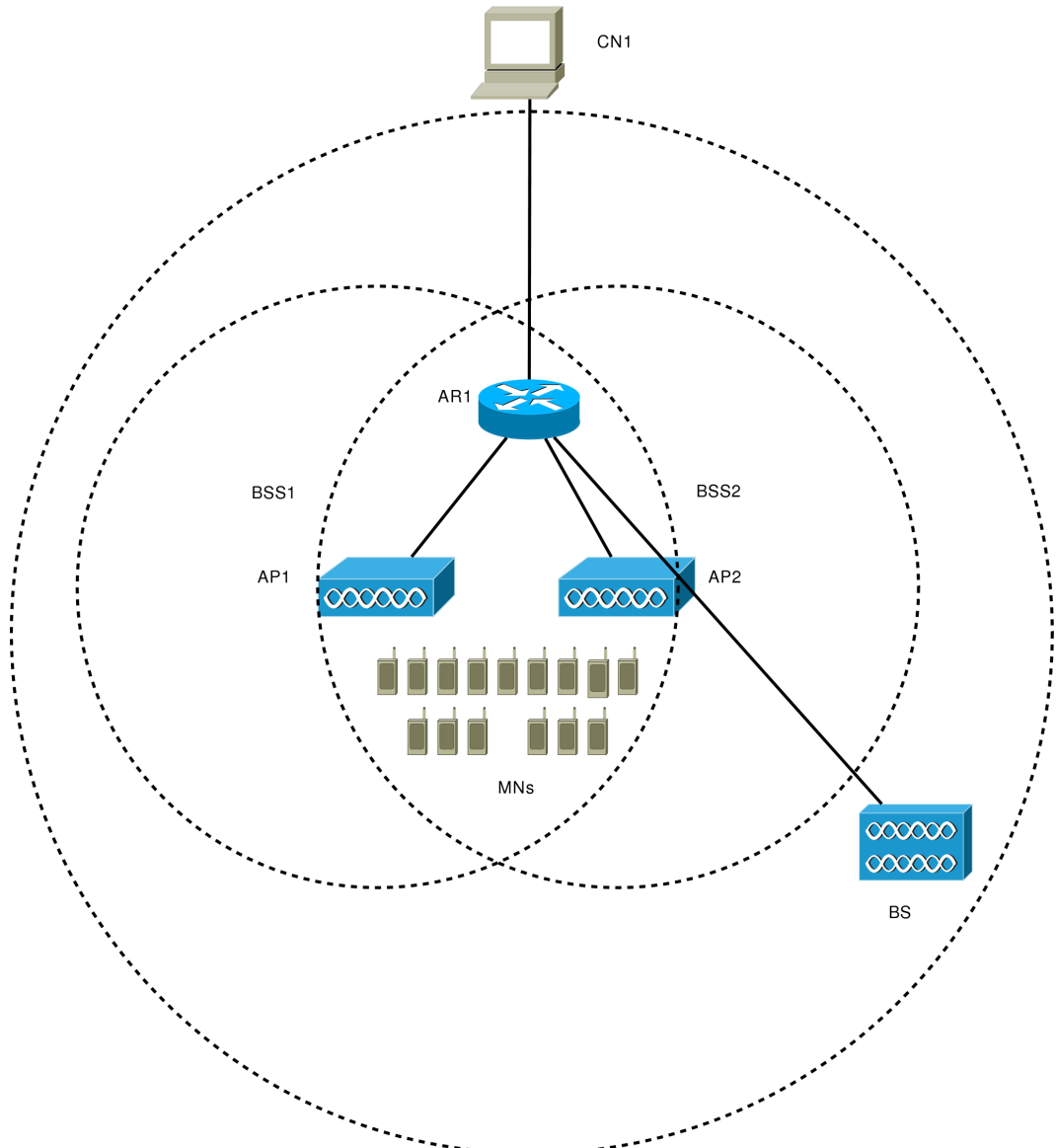

Figure 2. Experiment setup

The threshold ∆ as in Equation (1). is set at 250 kbps. $T_{ignore}$ is set to 1 second. $t_{repeat}$ is set to 200 ms and $n_{repeat}$ is set to 4. In cases IV, V and VI, MNs are allowed to ahndoff to BS if a suitable AP is not found to handoff. In case I, whenever the packet drop rate increases beyond 1% the MN suffering from performance degradation starts the process to handoff to another AP. In case II and case III, whenever the MN finds a rise in queue length beyond certain limit, say *qlength*, it starts the process of handoff. In case II *qlength* is set at 10 and in case III it is set at a very low value of 1.3. In case IV, whenever the packet drop rate increases beyond 1%, as in case I, the MN suffering from performance degradation starts the process to handoff to another AP. But unable to find a suitable AP, the MN performs handoff to BS. In case IV and case VI, whenever the MN finds a rise in queue length beyond certain limit, say *qlength*, it starts the process of handoff. In bothe these cases, the MN is unable to find a suitable AP to handoff and performs handoff to BS. In case V *qlength* is set at 10 and in case VI it is set at a very low value of 1.3 as in case of II and III respectively.

## 5.2 Results and Discussions

The results of our experiments are presented from Figure 3 to Figure 14. Figure 3 to Figure 8, gives the load vs throughput for AP1, AP2 and the whole ESS for the cases I, to VI respectively. Figure 9 to Figure 14 gives the delay incurred by different packets in AP1 and AP2 for case I to VI respectively. The total ESS throughput without using our scheme is put in each of Figure 3 to Figure 8 for ease of comparison. From the results we observe that when packet drop is used to sense performance degradation, the detection occurs late and when adjustment of load is done, the delay remains quite high. The throughput of the ESS drops a little and it takes some time for the throughput to recover. The detection occurs early when EWMA of queue length is used to detect performance degradation. An early adjustment avoids the loss of throughput. In all the cases, the network capacity increases when our scheme is used. In cases I, II and III, when the MNs are not allowed to handoff to BS, AP1 and AP2 gets satuarated when the number of MNs associated to the APs exceeds six. Throughput suffers and delay starts to increase. When MNs are allowed to handoff to BS, the APs no longer gets satuarated. In this cases, upto three MNs perform handoff to the BS releaving the APs. The delay starts to decrease. The delay is high when adjustment is done late as in case I , II, IV and V. For significant improvement in delay, sensing should be very early as in case III and VI. The sensing should be such that it detects a very small rise in the EWMA value of the queue length  an adjustment in load is done. The possible explanation for this is, if the EWMA of queue length rises considerably, the queue length is already high and even after the time the load is adjusted, packets take some significant time to pass through the queue. If the load is not excessively high, the queue length eventually decays. At high load, which is higher than the case if detection and adjustment is earlier like case III, the queue length decays slowly and the delay remains high during this period. In case II and V, the threshold value of EWMA of queue length is set at 10 and no significant reduction in delay is observed. In case III and VI, the threshold is set at a low value of 1.3 and the detection and adjustment is sufficiently early and there is significant improvement in delay. The best throughputs are observed when number of MNs per AP is within six. The lowest delays are observed when number of MNs per APs are within five.

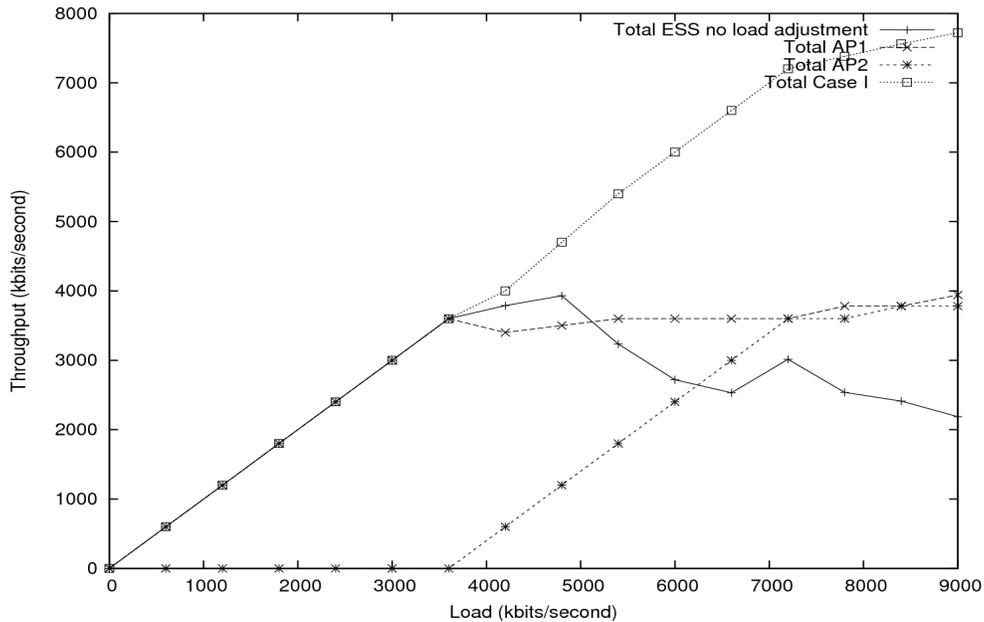

Figure 3. Throughput vs load :Case I

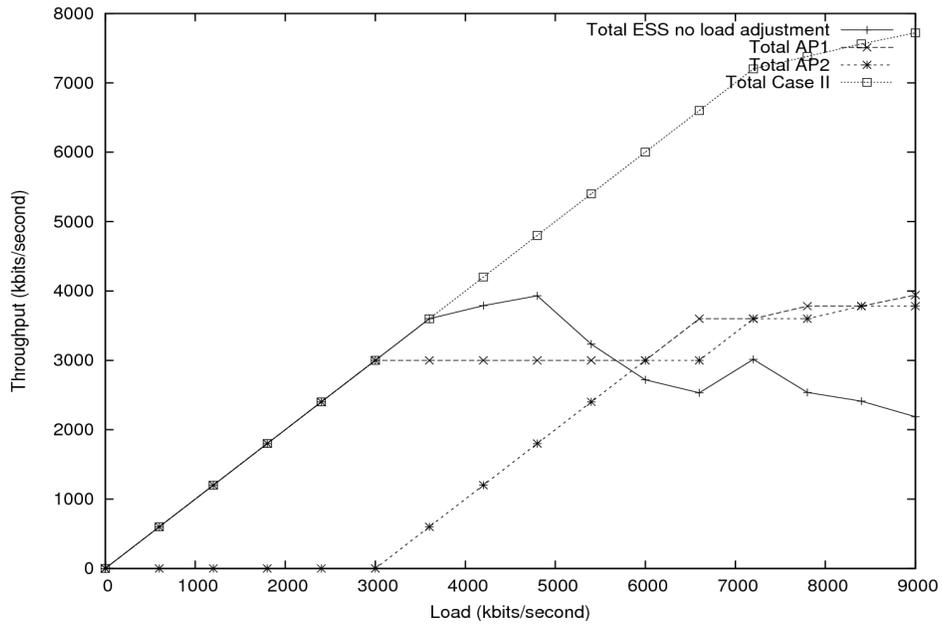

Figure 4. Throughput vs load :Case II

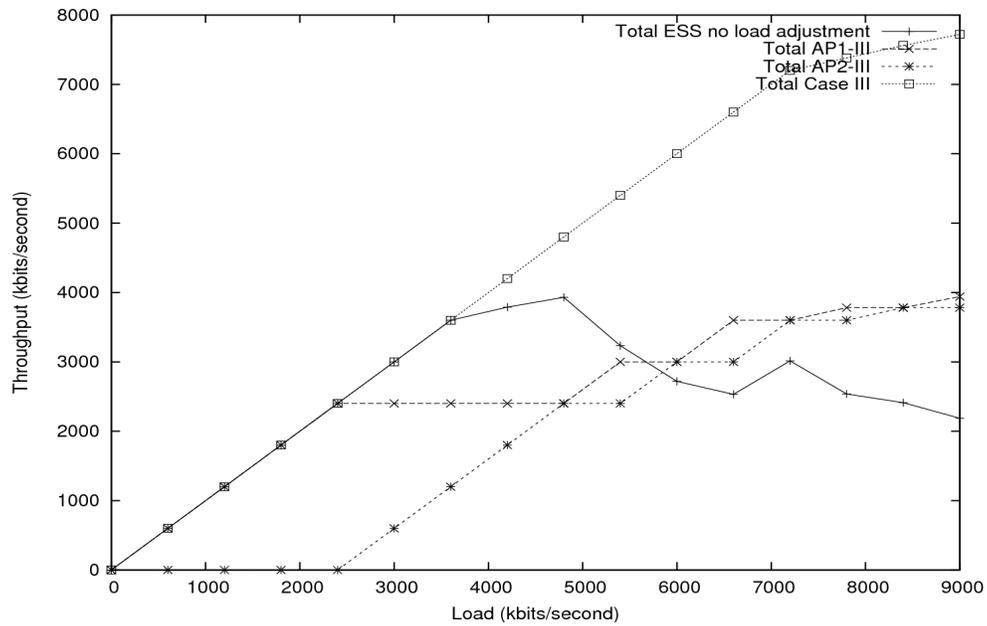

Figure 5. Throughput vs load :Case III

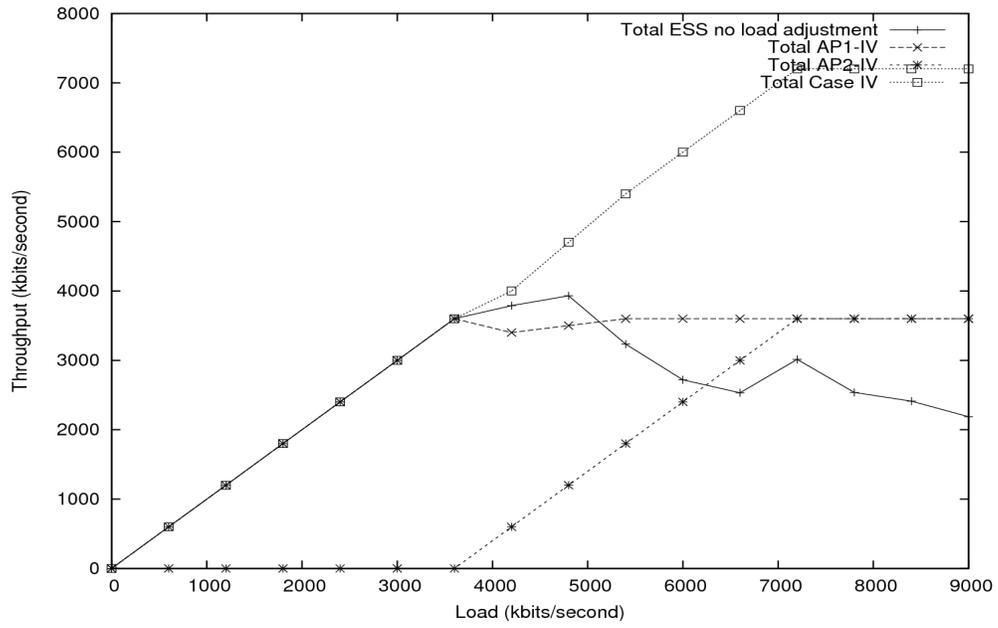

Figure 6. Throughput vs load :Case IV

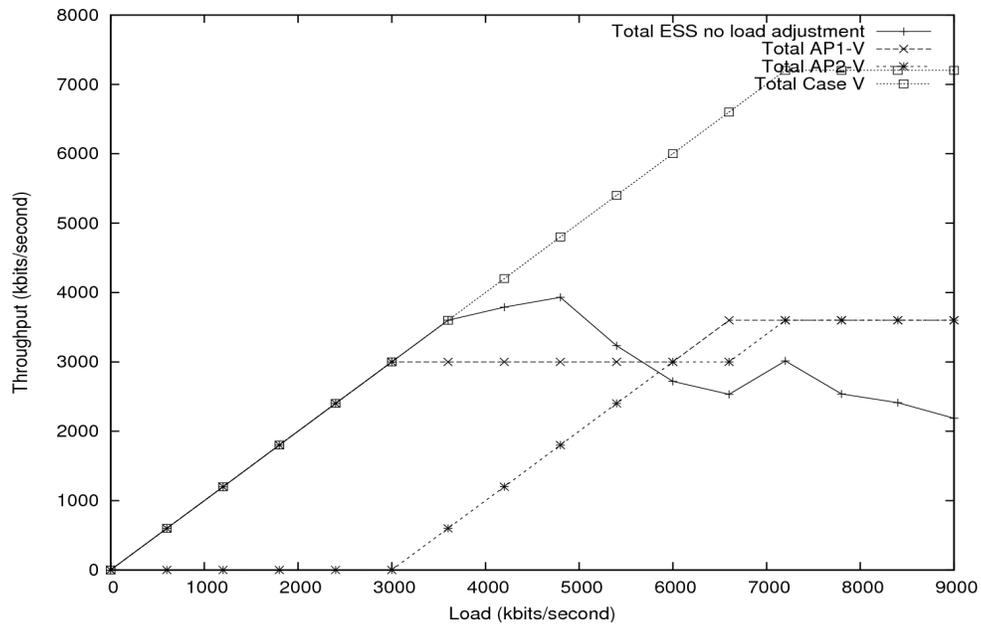

Figuer 7. Throughput vs load :Case V

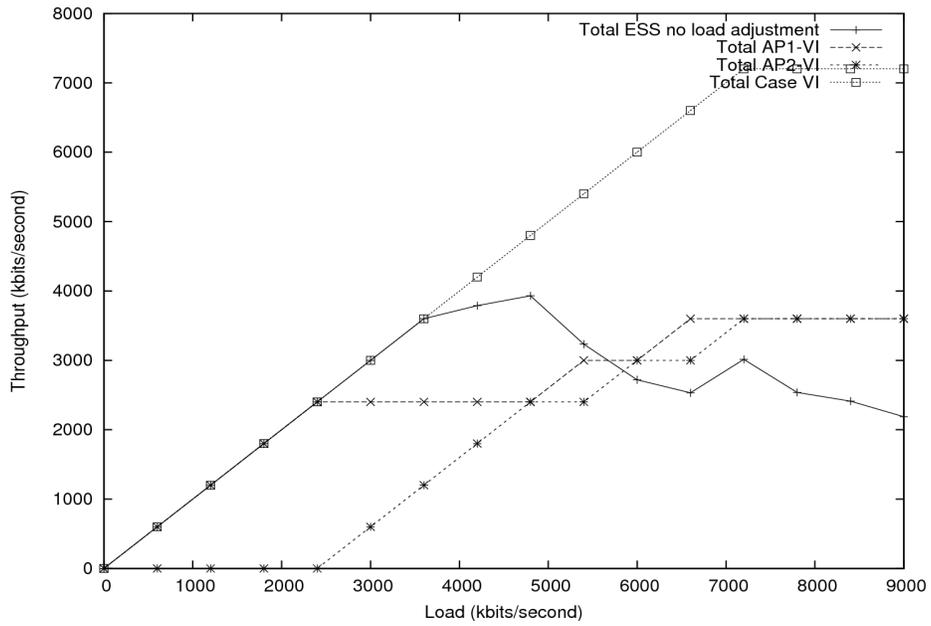

Figuer 8. Throughput vs load :Case VI

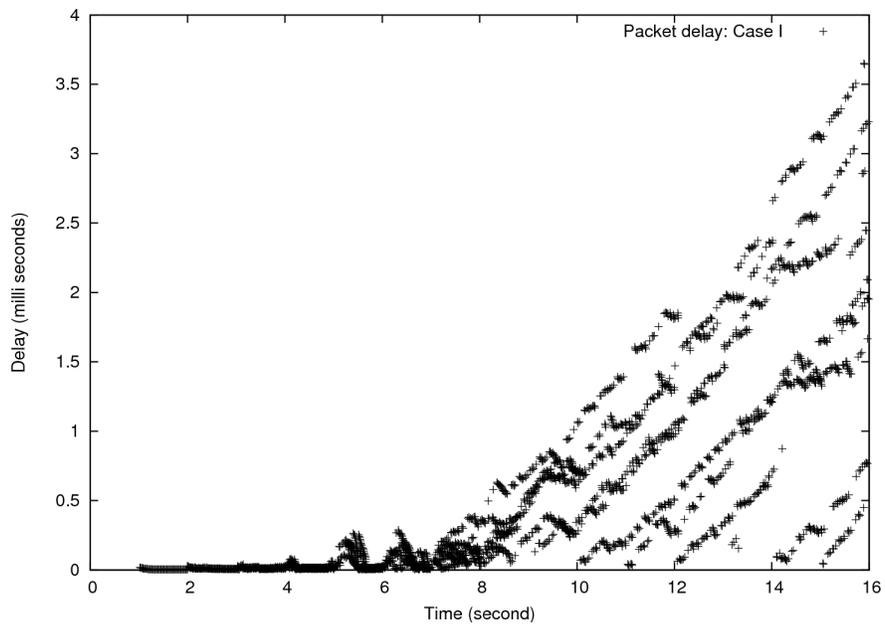

Figure 9. Delay characteristics :Case I

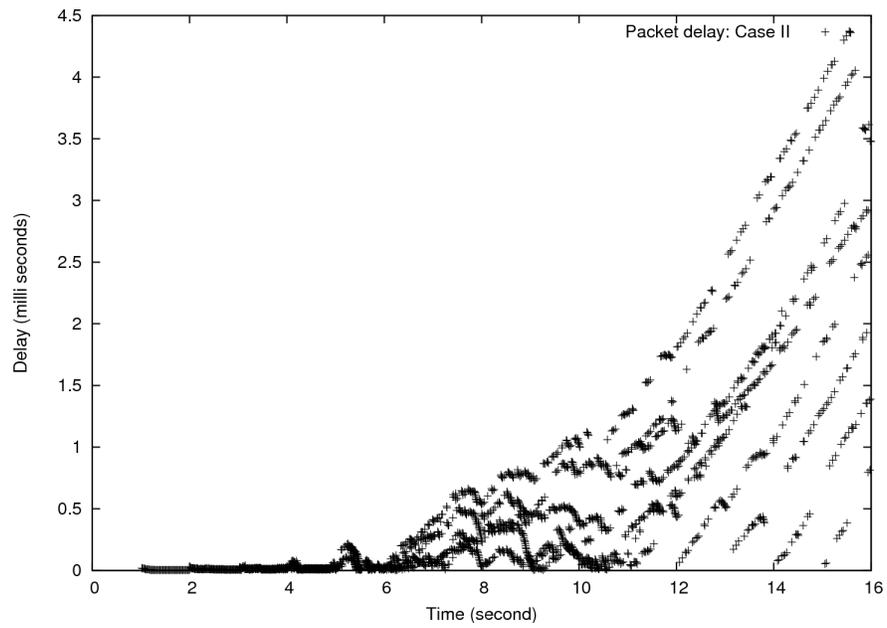

Figure 10. Delay characteristics :Case II

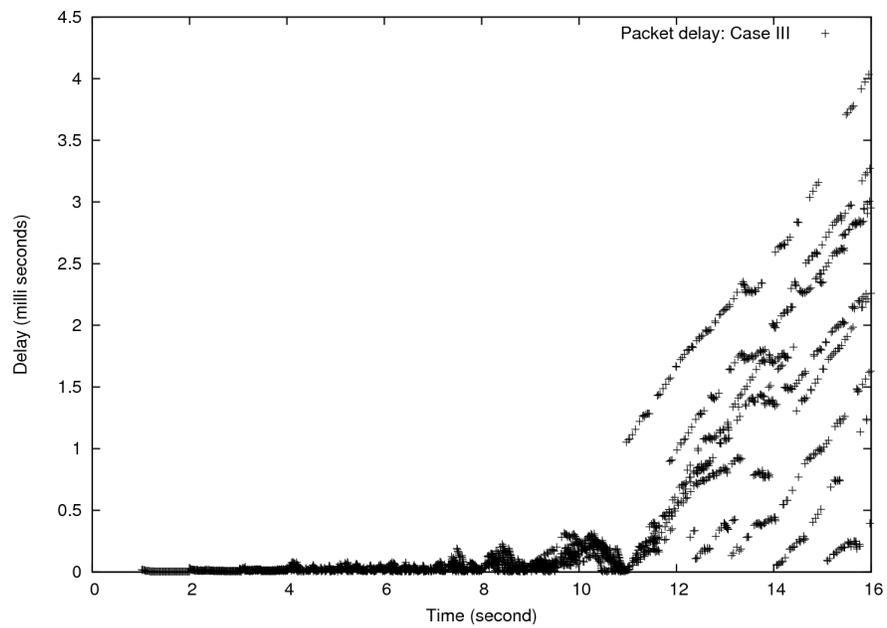

Figure 11. Delay characteristics :Case III

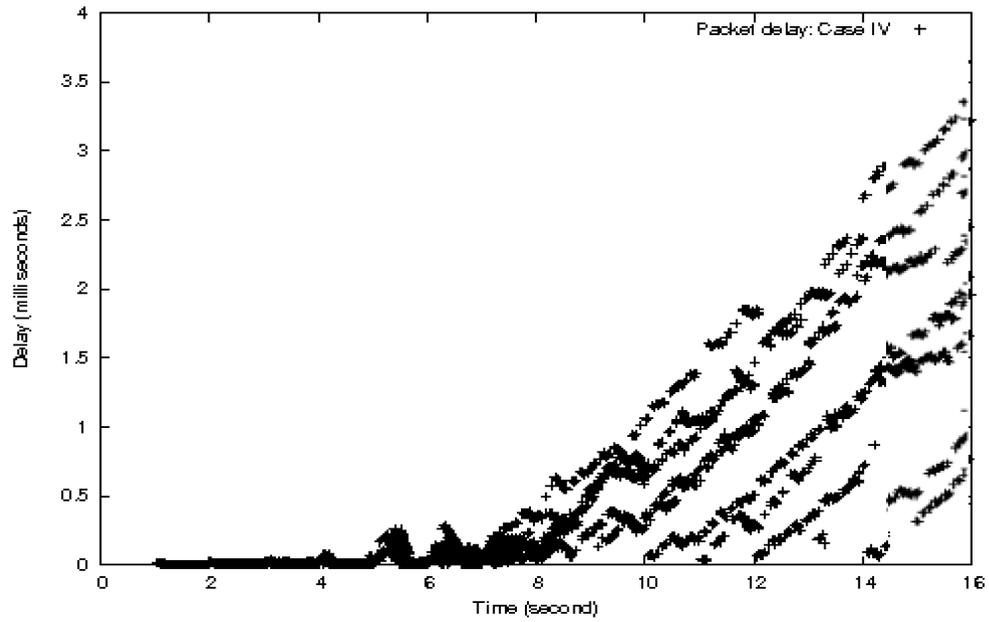

Figure 12. Delay characteristics :Case IV

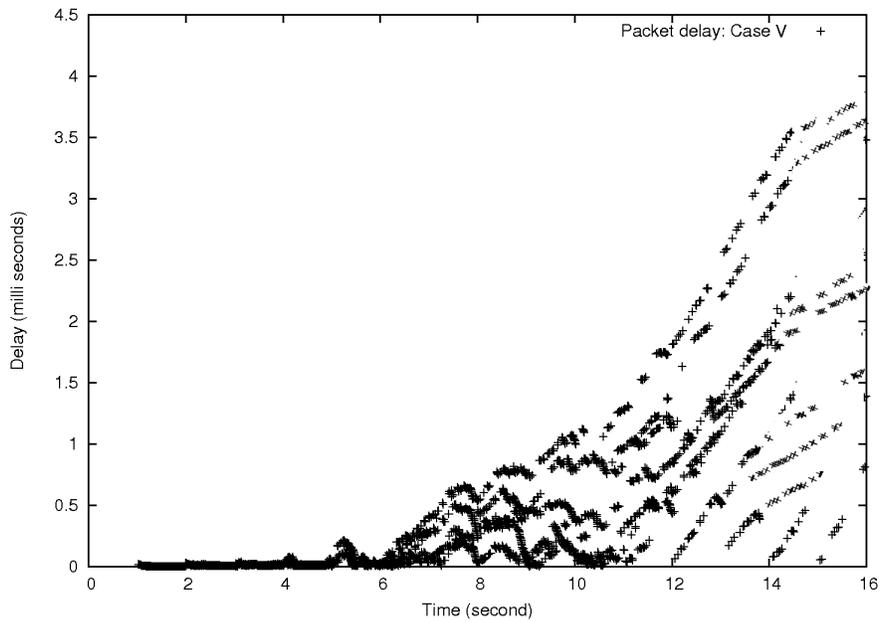

Figure 13. Delay characteristics :Case V

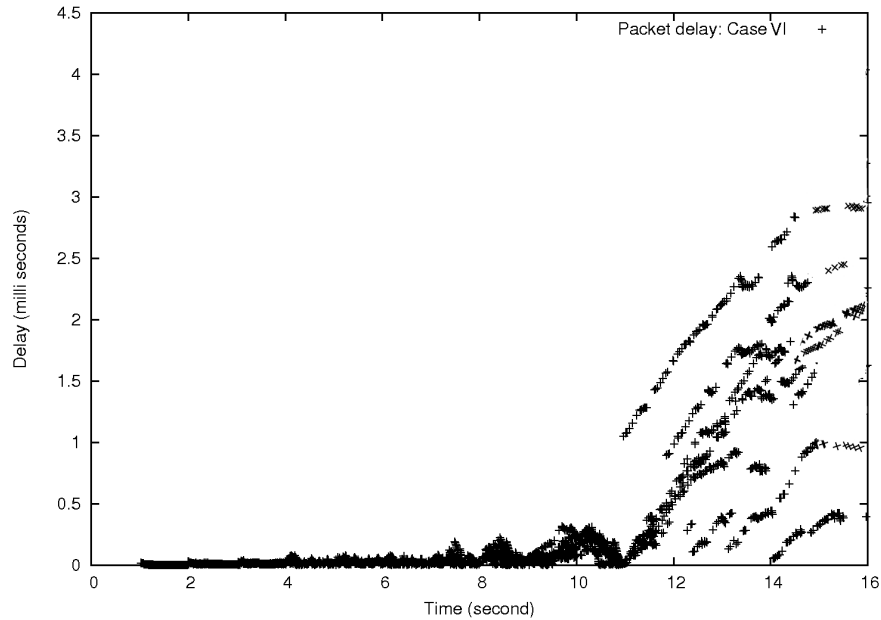

Figure 14. Delay characteristics :Case VI

When total number of MNs increases beyond twelve, throughput and delay of the whole ESS suffers as the ESS can no longer take the load without loss of performance. The best performance in terms of both throughput and delay is observed in case VI when the EWMA used to sense overload is set low and MNs are allowed to handoff to BS. The handoff latency in all cases are observed to be within 50 ms.

## 6. CONCLUSIONS

In this paper we have proposed a scheme to adjust the load of Aps in a WLAN whenever the MNs in some APs suffer from degradation of performance while other APs have spare capacity. By so adjusting the load, the performance as seen from the MNs as well as the overall performance of the WLAN increase. The scheme also transfers MNs to a BS if a suitable AP to continue communication without performance degradation cannot be found. The performance increase occurs in terms of both throughput and delay. Instead of a complete load balancing, performance improvement is done only by adjusting association of one MN at a time. A new scanning technique is proposed along with a new method to quickly update routing informations in access routers to reduce handoff lateny. The proposed scheme is tested by simulation. The results show clear performance improvement of an ESS in terms of throughput and delay.

International Journal of Computer Networks & Communications (IJCNC) Vol.3, No.3, May 2011Proceedings of the 2nd ACM international workshop on Wireless mobile applications and services on WLAN hotspots, pp. 51–60, 2004.

[4] M. Balazinska and P. Castro, "Characterizing mobility and network usage in a corporate wireless local-area network," MobiSys '03: Proceedings of the 1st international conference on Mobile systems, applications and services, pp. 303–316, 2003.

[5] A. Balachandran, G. M. Voelker, P. Bahl, and P. V. Rangan, "Characterizing user behavior and network performance in a public wireless LAN," Marina Del Rey, pp. 195–205, 2002.

[6] T. Henderson, D. Kotz, and I. Abyzov, "The changing usage of a mature campus-wide wireless network," Comput. Netw., vol. 52, no. 14, pp. 2690–2712, 2008.

[7] H. Zhai, X. Chen, and Y. Fang, "How well can the IEEE 802.11 wireless LAN support quality of service?" IEEE Transaction on Wireless Communications, vol. 4, pp. 3084–3094, 2005.

[8] 802.16-2004 IEEE IEEE Standard for Local and metropolitan area networks Part 16: Air Interface for Fixed Broadband Wireless Access Systems, IEEE, Édition 2004.

[9] G. Sawma, I. Aib, R. Ben-El-Kezadri, and G. Pujolle, "ALBA: An autonomic load balancing algorithm for IEEE 802.11 wireless networks," IEEE/IFIP Network Operations and Management Symposium (NOMS), pp. 891–894, 2008.

[10] H. Gong and J. Kim, "Dynamic load balancing through association control of mobile users in WiFi networks," IEEE Transactions on Consumer Electronics, vol. 54, no. 2, pp. 342–348, 2008.

[11] E. H. Ong and J. Y. Khan, "An integrated load balancing scheme for future wireless networks," ISWPC'09: Proceedings of the 4th international conference on Wireless pervasive computing, pp. 103–108, 2009.

[12] S. Sheu and C. Wu, "Dynamic load balance algorithm (DLBA) for IEEE 802.11 wireless LAN," Journal of Science and Engineering, vol. 2, 1999.

[13] IEEE Std 802.11k-2008, Part 11: Wireless LAN Medium Access Control (MAC) and Physical Layer (PHY) Specifications Amendment 1: Radio Resource Measurement of Wireless LANs, IEEE, Édition 2008.

[14] Y. Lee, K. Kim, and Y. Choi, "Optimization of AP placement and channel assignment in wireless LANs," LCN '02: Proceedings of the 27th Annual IEEE Conference on Local Computer Networks, pp. 831–836, 2002.

[15] Y. Bejerano, S. Han, and L. Li, "Fairness and load balancing in wireless LANs using association control," MobiCom '04: Proceedings of the 10th annual international conference on Mobile computing and networking, pp. 315–329, 2004.

[16] Y. Bejerano and S. Han, "Cell breathing techniques for load balancing in wireless LANs," IEEE Transactions on Mobile Computing, vol. 8, no. 6, pp. 735–749, 2009.

[17] R. Daher and D. Tavangarian, "Resource reservation and admission control in IEEE 802.11 WLANs," QShine '06: Proceedings of the 3rd international conference on Quality of service in heterogeneous wired/wireless networks, article 4, 2006.

[18] A. Balachandran, P. Bahl, and G. M. Voelker, "Hot-spot congestion relief in public-area wireless networks," WMCSA '02: Proceedings of the Fourth IEEE Workshop on Mobile Computing Systems and Applications, pp. 70–80, 2002.

[19] L. Du, M. R. Jeong, A. Yamada, Y. Bai, and L. Chen, "QoS aware access point selection for pre-load-balancing in multi-BSSs WLAN," IEEE Wireless Communications and Networking Conference(WCNC), pp. 1634–1638, 2008.
64

## Authors

**Abhijit Sarma** obtained B.E.(Electrical) degree from Jorhat Engineering College in 1984. He obtained M.C.A. Degree from Jorhat Engineering in 1990. Worked as Assistant Engineer and then Assistant Executive Engineer from 1987 to 1993 in Assam State Electricity Board.. Worked as Assiatant Professor in Gauhati University from 1993. He is currently doing PhD. In Indian Institute of Technology, Guwahati.

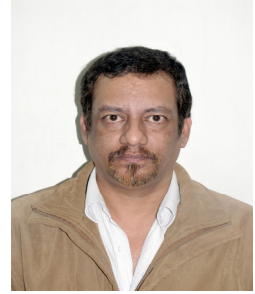

**Shantanu Joshi** Obtained his M.Tech degree from Indian Institute of Technology, Guwahati in 2010.

**Sukumar Nandi** received BSc (Physics), BTech and MTech from CalcuttaUniversity in 1984, 1987 and 1989 respectively. He received the PhD degree in Computer Science and Engineering from Indian Institute of Technology Kharagpur in 1995. In 1989–1990 he was a faculty in Birla Institute of Technology, Mesra, Ranchi, India. During 1991–1995, he was a scientific officer in Computer Science and Engineering, Indian Institute of Technology Kharagpur. In 1995 he joined Indian Institute of Technology Guwahati as an Assistant Professor in Computer Science and Engineering. Subsequently, he became Associate Professor in 1998 and Professor in 2002. He was in School of Computer Engineering, Nanyang Technological University, Singapore as Visiting Senior Fellow for one year (2002–2003). He was member of Board of Governor, Indian Institute of Technology Guwahati for 2005 and 2006. He was General Vice-Chair of 8th International Conference on Distributed Computing and Networking 2006. He was General Co-Chair of the 15th International Conference on Advance Computing and Communication 2007. He is also involved in several international conferences as member of advisory board/ Technical Programme Committee. He is reviewer of several international journals and conferences. He is co-author of a book titled "Theory and Application of Cellular Automata" published by IEEE Computer Society. He has published more than 150 Journals/Conferences papers. His research interests are Computer Networks (Traffic Engineering, Wireless Networks), Computer and Network security and Data mining. He is Senior Member of IEEE and Fellow of the Institution of Engineers (India)

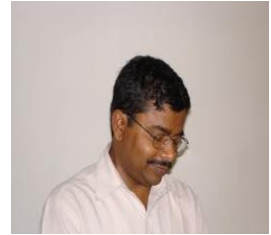